\documentclass[a4paper,11pt]{article}
\usepackage{pos}
\usepackage{amsmath}
\usepackage{amsfonts}
\usepackage{hyperref}
\usepackage{mathtools}
\usepackage{amsmath,epsfig}

\newcommand{\x}{\hat{x}}
\newcommand{\xih}{\hat{\xi}}
\newcommand{\Q}{\hat{Q}}
\newcommand{\Qb}{\hat{\bar{Q}}}

\title{Noncommutative spaces and superspaces from Snyder and Yang type models}

\author*[a]{Jerzy Lukierski}
\author[a]{Mariusz Woronowicz}

\affiliation[a]{Institute for Theoretical Physics, University of Wroclaw,\\
pl. Maxa Borna 9, 50-205 Wroclaw, Poland}

\emailAdd{jerzy.lukierski@uwr.edu.pl}
\emailAdd{woronow.m@gmail.com}

\abstract{The relativistic $D=4$ Snyder model is formulated in terms of $D=4$ $dS$ algebra $o(4,1)$ generators, with noncommutative Lorentz-invariant Snyder quantum space-time provided by $\frac{O(4,1)}{O(3,1)}$ coset generators. Analogously, in relativistic $D=4$ Yang models the quantum-deformed relativistic phase space is described by the algebras of coset generators $\frac{O(5,1)}{O(3,1)}$ or $\frac{O(4,2)}{O(3,1)}$. We extend these algebraic considerations by using respective $dS$ superalgebras, which provide Lorentz-covariant quantum superspaces (SUSY Snyder model) as well as relativistic quantum phase super spaces (SUSY Yang model).}

\FullConference{%
  Corfu Summer Institute 2021 "School and Workshops on Elementary Particle Physics and Gravity"\\
  29 August - 9 October 2021\\
  Corfu, Greece
}

\begin{document}
\maketitle

\section{Introduction}

\subsection{Conceptual problems with QFT approach to QG}

Recent efforts to describe on quantum level all fundamental interactions, with quantum gravity (QG) included, suggest that for such a purpose we should change the classical concept of commutative continuous space-time and canonical quantum phase spaces as well as modify the classical Lie-algebraic symmetries. The historical development provided various arguments that quantization of gravity requires introduction of noncommutative (NC) quantum space-time as well as quantum-deformed noncanonical phase space coordinates. Such nonclassical structures of quantum space -times follows from the dynamical role of space-time in general relativity (GR). One can argue (see e.g. \cite{1},\cite{2}) that due to gravitational Einsteinian dynamics it is not possible to observe effectively the distances smaller than the Planck length $\lambda_P\simeq 10^{-33}$cm, i.e $\lambda_P$ defines an absolute resolution limit in measurements of space-time distances. NC structure of quantum space-time can be linked with gravitational creation of microscopic black holes,  effectively replacing the notion of space-time points known from classical theories.

Already in 1935 M. Bronstein wrote \cite{3} \textit{"We can not localize arbitrarily large amount of mass or energy in very small volume - we will be not able to observe it because the action of gravitational forces will make such an observation  impossible."}

The gravitational forces due to quantum-mechanical effects represented by Heisenberg uncertainty relations will cause during the localization measurement the atomization of space-time, what leads effectively to quantum space-time described by NC operators
$$
\begin{array}{ccc}
\text{reaction of}  & & \text{below } 10^{-33} \text{ cm} \\
\text{dynamical space-time to} & \Leftrightarrow & \text{the notion of classical space-time}\\
\text{quantum measurement process} & & \text{looses its operational meaning.}
\end{array}
$$

Since almost hundred years it is well-known that in QM the limits of measurability of positions and momenta are determined by the NC structure of quantum-mechanical phase space algebra. In particular, the QG effects introduce changes of geometric space-time structure and leads to NC algebras for quantum space-time coordinates (see e.g. \cite{4}-\cite{6}). 

One can introduce in relativistic physics three basic dynamical frameworks, which are reflected in three different structure of phase space geometries:

\begin{enumerate}
\item
In classical relativistic theories $(\hslash=0,c\neq 0)$ i.e classical mechanics and classical field theory, the space-time $(x_\mu)$ and relativistic  phase space coordinates $(x_\mu,p_\mu)$ are classical, and we should use commutative geometric framework.

\item
One can consider relativistic quantum theories, but without QG interactions ($\hslash\neq 0,c\neq 0$, $G=0$, where $G$ defines the Newton constant describing gravity coupling). In such a way we arrive at relativistic quantum field theories (RQFT)\footnote{Unfortunately, the present status of relativistic quantum mechanics describing many interacting relativistic particles is not satisfactory.}, however without dynamically coupled gravitational fields. In such a case we deal with commutative space-times, but quantum-mechanical Heisenberg algebra which describes noncommutativity of coordinates and momenta leads the canonical NC structure of quantum-mechanical phase space.

\item
If we consider relativistic quantum field theories with coupled QG ($\hslash\neq 0, c\neq 0, G\neq0$) we study in principle the dynamical scheme which describes our present physical understanding of Universe. Besides NC space-time $\hat{x}_\mu$, with QG corrections taken into consideration, we arrive in this third geometric framework also at the algebra of noncanonical NC phase spaces, described by the NC basis ${(\hat{x}_\mu,\hat{p}_\mu})$\footnote{One gets the NC momenta already without QG effects, e.g. for particle moving in a gauge field background described by covariant gauge derivatives (e.g. in EM field background)}.
\end{enumerate}

One can describe schematically the passage from second to third framework as follows
$$
\begin{array}{ccc}
\text{classical}  & \xRightarrow{QG} & \text{NC quantum}\\
\text{space-time } x_\mu &  & \text{space-time } \hat{x}_\mu \\
 & & \\
\text{canonical NC}  & \xRightarrow{QG} & \text{QG-deformed noncanonical}\\
\text{phase space} & & \text{quantum phase space} 
\end{array}
$$
In field-theoretic formalism the passage from second to third dynamical framework is quite difficult because the quantized geometrodynamics described by QG can not be incorporated into the standard framework of QFT, with quantum fields defined on static (quite often flat) space-time manifolds. The standard quantum free fields are schematically described as a sum (integral) of quantized field oscillators $\hat{a}(p)$, with classical Fourier exponentials as linear coefficients
\begin{equation}
\hat{\phi}(x)=\int \hat{a}(p)e^{ipx}.
\label{aa1}
\end{equation}
However, due to QG modification of classical space-time the space-time coordinates $x_\mu$ appearing in (\ref{aa1}) become quantized, i.e we should replace in (\ref{aa1}) $x_\mu\rightarrow\hat{x}_\mu$. Because for quantum space-times we get 
\begin{equation}
[\hat{x}_\mu,\hat{x}_\nu]\neq 0
\label{aa2}
\end{equation}
in the presence of QG one can not work effectively with the formalism of standard quantum fields.

\subsection{Quantum space-times - three $D=4$ examples}
First task which one should consider is the description of possibly physically motivated examples of quantum space-times and quantum-deformed relativistic phase spaces. We list below three the most popular models of NC quantum space-times, which has been also extended to the respective quantum deformed relativistic phase spaces

{\bf{1. Snyder and Yang models}}\\
Snyder model \cite{8} was proposed still before invention of the renormalization procedure in 50's, in order to provide the regularization of infinities in QFT. In 60's, however, the perturbative renormalization theory has been established and only recently the interest in Snyder model has been revived due to other reasons in particular studying QG models. It provides an important example of Lorentz-covariant quantum space-time with Hermitean (real) NC coordinates $\hat{x}_\mu$, which are described by the coset generators of classical Lie algebra. If we decompose $D=4$ dS $o(4,1)$ algebra generators $\hat{M}_{ab}$ into $D=4$ Lorentz algebra and $\frac{O(4,1)}{O(3,1)}$ coset generators $(a,b=0,1,2,3,4;\mu=0,1,2,3)$ 
\begin{equation}
\hat{M}_{ab}=(\hat{M}_{\mu\nu},\hat{M}_{\mu4})
\label{aa3}
\end{equation}
after the identification
\begin{equation}
\hat{M}_{\mu4}\equiv\hat{x}_\mu
\label{aa4}
\end{equation}
one obtains the basic Snyder relation, describing space-time noncommutativity at ultrashort distances
\begin{equation}
[\hat{x}_\mu,\hat{x}_\nu]=il_P^2\hat{M}_{\mu\nu}
\label{aa5}
\end{equation}
where $l_P\simeq 10^{-33}$ cm denotes the Planck length.

Following old ideas of Dirac and Zeldovich \cite{9},\cite{10}, who advocated structural link between cosmological and microworld physics, if we use the quantum version of Born duality relation $(\hat{x}_\mu\leftrightarrow\hat{p}_\mu)$ \cite{11}, after supplementing $l_P\leftrightarrow\frac{1}{R^2}$ and identification (\ref{aa4}) one obtains standard $D=4$ de-Sitter algebra describing space-time at cosmological distances
\begin{equation}
[\hat{p}_\mu,\hat{p}_\nu]=i\frac{1}{R^2}\hat{M}_{\mu\nu}
\label{aa6}
\end{equation}
where NC generators $\hat{p}_\mu$ describe curved dS space-time translations and $R\simeq 10^{29}$ cm describes the radius of the Universe.

Almost simultaneously, in 1947 \cite{12}, Yang extended Snyder model in order to obtain $D=4$ quantum-deformed relativistic phase space $(\hat{x}_\mu,\hat{p}_\mu)$, with NC coordinates described by  (\ref{aa5})-(\ref{aa6}). In Yang model one uses the coset generators of $D=5$ dS Lie algebra, with $o(5,1)$ algebra decomposed into $D=4$ Lorentz algebra and the $\frac{O(5,1)}{O(3,1)}$ coset generators which provide the quantum-deformed relativistic phase space algebra. We decompose $o(5,1)$ as follows
\begin{equation}
\hat{M}_{AB}=(\hat{M}_{\mu\nu},\hat{M}_{\mu4},\hat{M}_{\mu5},\hat{M}_{45})
\label{aa7}
\end{equation}
where $A,B=0,1,\dots,5;$ besides (\ref{aa4}) we postulate 
\begin{equation}
\hat{M}_{\mu5}\equiv\hat{p}_\mu,\qquad \hat{M}_{45}\equiv \hat{h}
\label{aa8}
\end{equation}
where $(\hat{x}_\mu,\hat{p}_\mu,\hat{h})$ represent the Lorentz-covariant quantum-deformed Heisenberg algebra. Its NC structure is defined by relations (\ref{aa5})-(\ref{aa6}) and
\begin{equation}
[\hat{x}_\mu,\hat{p}_\nu]=i\eta_{\mu\nu}\hat{h}
\label{aa8a}
\end{equation}
where $\eta_{\mu\nu}=(-1,1,1,1)$ denotes $D=4$ Lorentz metric.

It should be stresses that the construction presented above implies automatically the Lorentz covariance of both noncommutative Snyder and Yang models.

{\bf{2. $\kappa$-Minkowski NC space}}\\
In late 80-ties there was discussed the question how to introduce the quantum deformation of Poincare algebra which incorporates geometrically, besides $c$ and $\hslash$, the third fundamental constant, described by Planck mass $m_P$ or Planck length $\lambda_P$ ($\lambda_P=\frac{\hslash}{m_Pc}$; if we put $\hslash=c=1$ one gets $m_P=\frac{1}{\lambda_P}$). The answer was provided in 1991 by $\kappa$-deformation of Poincare-Hopf algebra \cite{13}, an example of quantum symmetry algebras which due to Drinfeld \cite{14} are described by noncocommutative Hopf algebras and called quantum algebras. The $\kappa$-deformed Minkowski space-time can be obtained as NC representation (quantum module) of $\kappa$-deformed quantum Poincare algebra \cite{15},\cite{16}. In its historically first standard form the NC structure of Poincare algebra looks as follows\footnote{If $a_\mu=(1,0,0,0)$ the formula (\ref{aa10}) is a special case of generalized $\kappa$-Minkowski spaces, parametrized by constant fourvector $a_\mu$ as follows \cite{17}
\begin{equation}
[\hat{x}_\mu,\hat{x}_\nu]=i\frac{\hslash}{\kappa c}(a_\mu\hat{x}_\nu-a_\nu\hat{x}_\mu).
\label{aa11}
\end{equation}
The fourvector $a_\mu$ determines the $\kappa$-deformed quantum direction in Minkowski space-time (in formula (\ref{aa10}) there is quantum-deformed the time direction.}
\begin{equation}
[\hat{x}_0,\hat{x}_i]=i\frac{\hslash}{\kappa c}\hat{x}_i,\qquad[\hat{x}_i,\hat{x}_j]=0
\label{aa10}
\end{equation}
where in physical assignments to elementary particle physics we put $\kappa=m_P$.

Hopf-algebraic framework introduces dual pairs ($H,\tilde{H}$) of Hopf algebras, in $\kappa$-Poincare case $H$ describing $\kappa$-deformed Poincare algebra ($\kappa$-Poincare algebra) and $\tilde{H}$ $\kappa$-deformation of Poincare matrix group ($\kappa$-Poincare group). The $\kappa$-deformed Minkowski space can be described by the deformed translational sector of $\kappa$-Poincare group. From dual pair of $\kappa$-Poincare algebra and $\kappa$-Poincare group one can construct the Heisenberg double algebra \cite{18},\cite{19}, which describes the generalization of $\kappa$-deformed quantum phase space \cite{20}.

{\bf{3. $\theta_{\mu\nu}$-deformed Minkowski space (DSR model)}}\\
The most popular and very simple quantum deformations of relativistic Minkowski space are parametrized by constant antisymmetric numerical tensor $\theta_{\mu\nu}=-\theta_{\nu\mu}$, which leads to the following commutator of NC quantum space-time
\begin{equation}
[\hat{x}_\mu,\hat{x}_\nu]=il^2\theta_{\mu\nu}
\label{aa12}
\end{equation}
where $l$ describes the elementary length, which may be also identified with Planck length $(l=l_P)$.

Such a model was introduced firstly as NC quantum space, but later it was obtained as the quantum translational sector of Poincare-Hopf algebra $H_\theta$, where $\theta\equiv\theta_{\mu\nu}$ describe six deformation parameters which break Lorentz covariance. In 1994-1995 (see \cite{1},\cite{2}) this model was considered in the context of QG by Dopplicher, Fredenhagen and Roberts, so often it is called DFR model of quantum Minkowski space.

The Hopf algebra $H_\theta$ belongs to simpler class of quantum groups, which is obtained by Abelian twist quantization of classical Poincare-Hopf algebra. In $H_\theta$ the Poincare algebra after twisting remains classical, but primitive coproducts are modified. Due to Hopf algebraic duality relations in $H_\theta$ there is deformed the coalgebra and in $\theta_{\mu\nu}$-deformed Poincare-Hopf group $\tilde{H}_\theta$ the algebra of space-time translations is becoming noncommutative (see (\ref{aa12})).

The advantage of twist quantization is provided by the presence of explicit formulae expressing the local products of NC fields $\phi_i(\hat{x})$ as nonlocal products of classical fields $\phi_i(x)$, with the use of nonlocal multiplication described by so-called star product ($\star$-product)
\begin{equation}
\phi_i(\hat{x})\cdot\phi_j(\hat{x})\xrightarrow{\text{Weyl map}}\phi_i(x)\star\phi_j(x).
\label{aa13}
\end{equation}
In such a way the algebra of twist-deformed quantum fields can be expressed in terms of standard fields on classical Minkowski space.
In the case of $\theta_{\mu\nu}$-deformation the $\star$-product (\ref{aa13}) has been introduced much earlier in statistical physics as nonlocal Moyal product $\star_M$ \cite{7}
\begin{equation}
[\hat{x}_\mu,\hat{x}_\nu]\simeq [x_\mu,x_\nu]_{\star_M}\equiv x_\mu\star_M x_\nu-x_\nu\star_M x_\mu=il^2\theta_{\mu\nu}.
\label{aa14}
\end{equation}

The Poincare algebra generators $(p_\mu, M_{\mu\nu})$ of $H_\theta$ are classical and the generators $(\hat{x}_\mu, \hat{\Lambda}_{\mu\nu})$ of $\tilde{H}_\theta$ are quantum-deformed. From $H_\theta$ and $\tilde{H}_\theta$ can be obtained the Heisenberg double algebra 
\begin{equation}
\mathcal{H}_\theta=\tilde{H}_\theta\rtimes H_\theta
\end{equation}
which provides the generalized $\theta_{\mu\nu}$-deformed phase space $\mathcal{P}_\theta^{10;10}=(\x_\mu,\hat{\Lambda}_{\mu\nu};p_\mu,M_{\mu\nu})$ and standard $\theta_{\mu\nu}$ deformed phase space $\mathcal{P}_\theta^{4;4}=(\x_\mu,p_\mu)$ described by relation (\ref{aa12}) and
\begin{equation}
[\x_\mu,p_\nu]=i\eta_{\mu\nu},\qquad[p_\mu,p_\nu]=0.
\label{aa15}
\end{equation}
It should be added that due to the appearance of $c$-number as value of commutator $[\x_\mu,\x_\nu]$ (see (\ref{aa12})), the standard $\mathcal{P}_\theta^{4;4}$ as well as generalized $\theta_{\mu\nu}$-deformed $\mathcal{P}_\theta^{10;10}$ covariant relativistic phase spaces are described by Poincare-Hopf algebroids (see e.g. \cite{21}).

\subsection{From quantum space-times to quantum deformed spinors and quantum superspaces}

In several well-known geometric frameworks in physics (e.g. Penrose twistor description or supersymmetric theories) basic coordinates are described by spinors or superspaces with both vectorial and spinorial coordinates. In this paper we investigate the constructions of quantum-deformed spinors as well as quantum-deformed superspaces.

In Snyder model the reinterpretation of $\frac{O(4,1)}{O(3,1)}$ coset generators as noncommutative coordinates provided the first example of Lorentz -covariant quantum space-time. Analogously, in Yang model the $\frac{O(5,1)}{O(3,1)}$ generators can be treated as the linear basis of relativistic $D=4$ quantum phase space algebra. These constructions one can generalize in two ways:
\begin{enumerate}
\item{In order to describe the quantum-deformed spinorial coordinates one should use the cosets of groups which are describing spinorial coverings $\overline{O(D-k,k)}$ of (pseudo)orthogonal space-time symmetry groups. Such cases provide bosonic quantum spinors which are obtained by deformation of commuting spinorial spaces.}
\item{One gets quantum-deformed fermionic spinors and quantum superspaces if we consider Lie supergroups and superalgebras, with its bosonic sector containing space-time symmetries (e.g. Lorentz symmetries)\cite{22}. By supergeneralization of relation (\ref{aa4}) one can identify the supercharges with fermionic sector of quantum superspace. We should observe that the quantum space-time for semi-simple superalgebras  due to basic superalgebra relations are expressed as composite algebraic objects which are bilinear in terms of quantum-deformed fermionic spinors. Because in semi-simple case the supercharges describe the algebraic basis of superalgebra, we obtain quantum superspaces containing quantum fermions as elementary spinor variables and composite quantum space-time coordinates.}
\end{enumerate}

Finally let us very briefly describe the content of this paper. In Sect. 2 we describe $D=4$ Lorentz-covariant Snyder and Yang models which provide Snyder quantum space-times and Yang quantum-deformed relativistic phase spaces. In Sect. 3 we provide the supersymmetric extension of $D=4$ Snyder quantum-deformed space-times. In Sect. 4 we consider the supersymmetric $D=4$ Yang model providing the supersymmetric quantum-deformed phase superspaces with composite bosonic sector. In last Sect. 5 we provide on outlook, in particular as novelty we indicate how one can introduce quantum-deformed Snyder models as obtained from twist-deformed quantum $D=4$ AdS or dS algebras.

\section{{Snyder quantum space-times and Yang quantum phase spaces}}

\subsection{Snyder dS and AdS quantum space-times}

D=4 dS and AdS algebras are described by the following five-dimensional pseudo-orthogonal algebras (A=0,1,2,3,4)
\begin{equation}
[M_{AB}, M_{CD}] = i (\eta_{AD} M_{BC} +\eta_{BC}M_{AD} - \eta_{AC}M_{BD} - \eta_{BD}M_{AC})
\label{wz0}
\end{equation}
  with signature $\eta_{AB} = diag(-1,1,1,1,\epsilon)$ and $\epsilon = \eta_{44}=\pm 1$, where $\epsilon=1$ for dS algebra and $\epsilon=-1$ for AdS algebra.
 Following (\ref{aa4}) we postulate that $M_{\mu 4} = \frac{1}{\lambda} \hat{x}_\mu$,
  where $\lambda$ is an elementary length, in physical applications often identified with the Planck length
   $\lambda_p = \sqrt{\frac{\hbar G}{c^3}} \simeq 1.6 \cdot 10^{-33}$cm. 
 We obtain besides the Lorentz algebra  generators $M_{\mu\nu}$  ($\mu=0,1,2,3)$  the following relations defining NC Snyder 
 space-times\footnote{In the paper we choose $\hslash=c=1$ with only length or mass dimensionalities taken into consideration.} and described by $\frac{O(4,1)}{O(3,1)}$ and $\frac{O(3,2)}{O(3,1)}$ cosets
 \begin{eqnarray}
 &
  [ M_{\mu\nu} , \hat{x}_{\rho} ] = i ( \eta_{\nu\rho} \hat{x}_{\mu} - \eta_{\mu \rho} \hat{x}_{\eta})
  \label{wz1}
 \\
 &
 \label{wz2}
  [ \hat{x} _{\mu} , \hat{x}_{\nu} ] = i \epsilon \beta M_{\mu \nu} \qquad \beta = \lambda^{2} > 0
  \label{wz3}
 \end{eqnarray}
 The difference between NC dS and AdS Snyder 
space-times consists only  in difference of sign on rhs of relation (\ref{wz2}).

The algebra with the basis ($M_{\mu\nu}, \hat{x}_\rho$) describes an elementary relativistic quantum object, namely D=4 quantum (A)dS space-times $\hat{x}_\mu$ as Lorentz algebra module and classical Lorentz transformations generated by $M_{\mu\nu}$ providing the D=4 relativistic covariance
  of  Snyder equations (\ref{wz1}-\ref{wz2}).
Snyder models  (see (\ref{wz0}-\ref{wz2})) are Born-dual ($\hat{x}_\mu \leftrightarrow \hat{p}_\mu$, $M_{\mu\nu}$ 
 unchanged,
$\lambda \to \frac{1}{R}$) to the momentum space realizations $(M_{\mu\nu},\hat{p}_\mu)$ of 
  $o(4,1)$ or $o(3,2)$ algebras, with their generators
   describing the automorphisms of five-dimensional pseudospheres $(\epsilon=\pm 1)$
   \begin{equation}
   x^A \eta_{AB} x^B = - x^2_0 + x^2_1 + x^2_2 +x^2_3 +\epsilon x^2_4 = R^2
   \label{wz23a}
   \end{equation}
    where
 $M_{\mu 4}= R\hat p_{\mu}$ are the NC generators of 
curved translations on the pseudospheres $\frac{O(4,1)}{O(3,1)}$ 
or $\frac{O(3,2)}{O(3,1)}$\footnote{Snyder construction was
 already in sixties applied to the description of de-Sitter Universe ($\epsilon=1$), with R describing the 
 cosmological de-Sitter radius (see e.g. \cite{23},\cite{24}).}.

In fact Snyder constructed his  model as aimed at the description of NC geometry at ultra short (Planckian) distances,
in order to regularize geometrically the ultraviolet divergencies of renormalization procedure for locally interacting quantized fields.
Born duality formalizes physical 
 as well as some philosophical 
 concept that one can relate in physics the micro and macro world  phenomena - the first ones due to
quantum nature described by NC geometry, and the second 
  linked with classical de-Sitter dynamics in general relativity
 at very large cosmological distances. 

\subsection{Yang $D=4$ quantum phase spaces}

Already in 1947 C.N. Yang \cite{12} observed that by considering D=6 rotations algebras (i.e. putting in (\ref{wz0})  $A=0,1,2,3,4,5$)
 one can interpret the presence of  rotation generators $M_{5\mu}$ as adding to Snyder model the NC
 fourmomenta $\hat{p} _\mu$. 
 The sixth dimension can be added
   to dS or AdS Snyder model  
  in two-fold way,  by postulating that $\eta_ {55} =\epsilon' = \pm 1$.
 Assuming that $M_{\mu 5}= R \hat{p} _{\mu}$ one gets the following extension of Snyder
 equations (\ref{wz1})-(\ref{wz2}):
 \begin{equation}
 [ M _{\mu\nu}, \hat{p}_{\rho} ] = i (\eta_{\nu\rho} \hat{p}_{\mu} - \eta_{\mu \rho} \hat{p}_{\mu})
 \end{equation}
\begin{equation}\label{2.6}
[ \hat{p}_\mu , \hat{p}_\nu ] = i \epsilon' \gamma  M_{\mu\nu} 
 \qquad   \gamma = \frac{1}{R^2}
\end{equation}
Additionally besides (\ref{wz3}) and (\ref{2.6}) one gets the quantum-deformed noncanonical phase space commutator
\begin{equation}
[ \hat{x}_{\mu} , \hat{p}_\nu ] = i \frac{\lambda}{R} \eta_{\mu\nu} M_{45} = i \eta_{\mu\nu} \hat{d}
\label{dfds}
\end{equation}
with operator-valued substitution of Planck constant $\hbar$ by rescaled dilatation generator $\hat{d} = \frac{\lambda}{R} M_{45}$, which is a $D=4$ Lorentz scalar
 $( [ M_{\mu\nu}, \hat{d} ] = 0)$. The generator  $M_{45}$ commutes with $\hat{x}_\mu$ and $\hat{p}_\mu$ as follows
 \begin{equation}
 [ \hat{d}, \hat{x}_\mu ] = i \epsilon\lambda R \hat{p}_\mu
 \label{l11}
 \end{equation}
 
 \begin{equation}
 [ \hat{d} , \hat{p}_\mu ] = i \frac{\epsilon'}{\lambda R} \hat{x} _\mu
 \label{l22}
 \end{equation}
 and describes internal symmetry generators $(o(2)$ if $\epsilon=-\epsilon'$ or $o(1,1)$ if $\epsilon=\epsilon'$).
  
 In such a way we can obtain four types of Yang models, with two dychotomic parameters
  $\epsilon = \pm 1$ (see(\ref{wz2}))  and $\epsilon' = \pm 1$ (see (\ref{2.6})), which could be called dS-dS, dS-AdS, AdS-dS, AdS-AdS Yang models.
  
The scheme presented above has been very recently generalized (see \cite{25}), where in particular it is generalized basic quantum phase space commutator (\ref{dfds}).

\subsection{Snyder quantum phase spaces}

$D=4$ Snyder model described in Sect.2.1 is defined by ten independent generators $(M_{\mu\nu},\hat{x}_\mu\equiv M_{4\mu})$ of $D=4$ de-Sitter $o(4,1)$ algebra, and in order to obtain the algebra of noncanonical relativistic quantum phase space one should enlarge the set of $\hat{o}(4,1)$ Lie algebra generators. In $D=4$ Yang model (see Sect. 2.2) the noncommutative fourmomentum generators $\hat{p}_\mu$ are obtained by adding the fifth space dimension to the generators of $D=4$ Snyder model, i.e we deal with $D=5$ de-Sitter algebra $o(5,1)$, where $\hat{p}_\mu=M_{5\mu}$. In order to obtain quantum phase space Snyder in \cite{8} did not use any group-theoretic techniques, but extended the algebra (\ref{wz0})-(\ref{wz3}) by supplementing the Lorentz-covariant classical fourmomenta $p_\mu$ satisfying the relations
\begin{eqnarray}
&&[M_{\mu\nu},p_\rho]=i(\eta_{\nu\rho}p_\mu-\eta_{\mu\rho}p_\nu)\label{wz2.10}\\
&&[p_\mu,p_\nu]=0.\label{wz2.11}
\end{eqnarray}
Further in \cite{8} there were added also the deformed Heisenberg algebra relation
\begin{equation}
[\hat{x}_\mu,p_\nu]=i(\eta_{\mu\nu}+\beta p_\mu p_\nu) \label{wz2.12}
\end{equation}
in a way consistent with Jacobi identities.

The relations (\ref{wz0})-(\ref{wz3}) and (\ref{wz2.10})-(\ref{wz2.12}) define Lorentz-covariant noncanonical Snyder quantum phase algebra, with the generators $M_{\mu\nu}$ and $\hat{x}_\mu$ which can be represented in terms of classical Heisenberg algebra basis $(x_\mu,p_\mu)$ \cite{26}
\begin{equation}
[x_\mu,p_\nu]=i\eta_{\mu\nu},\qquad [x_\mu,x_\nu]=[p_\mu,p_\nu]=0.\label{wz2.13}
\end{equation}
It should be stressed that in this paper we consider the generators $(\hat{x}_\mu,M_{\mu\nu})$ of Snyder model and $(\hat{x}_\mu, p_\mu, M_{\mu\nu}, \hat{d})$ of Yang model as algebraically independent, however in Snyder quantum phase algebra there is used the representation in which the Lorentz algebra generators can be realized in terms of classical phase space coordinates (\ref{wz2.13}) in standard way  
\begin{equation}
M_{\mu\nu}=i(x_\mu p_\nu-x_\nu p_\mu).
\end{equation}
The choice (\ref{wz2.12}) can be consistently extended as follows \cite{27},\cite{28}
\begin{equation}
[\hat{x}_\mu,p_\nu]=i\{\eta_{\mu\nu}(1+\beta p^2)-\beta' p_\mu p_\nu\} \label{wz2.15}
\end{equation}
however in such a case the relation (\ref{wz2}) should be modified in the following way \cite{29} (both parameters $\beta$ and $\beta'$ have the length square dimension)
\begin{equation}
 [ \hat{x} _{\mu} , \hat{x}_{\nu} ] = i \frac{(2\beta-\beta')+(2\beta+\beta')\xi}{1+\xi} M_{\mu \nu}.\label{wz2/16}
\end{equation}
The general modification of the relations (\ref{wz2.12})-(\ref{wz2.15}) given by the formula ($\xi=\eta p^2$ is dimensionless)
\begin{equation}
[\hat{x}_\mu,p_\nu]=i(F(\xi)\eta_{\mu\nu}1+\beta p_\mu p_\nu G(\xi)) \label{wz2.15a}
\end{equation}
require the following generalization of the formula (\ref{wz2/16})
\begin{equation}
 [ \hat{x} _{\mu} , \hat{x}_{\nu} ] = i \Psi(\xi)M_{\mu\nu}\label{wz2.16}
\end{equation}
where from the Jacobi identities it follows that \cite{26},\cite{30}
\begin{equation}
\Psi(\xi)=F(\xi)G(\xi)-2(F(\xi)+\xi G(\xi))\frac{dF}{d\xi}.
\end{equation}
The Snyder choice (\ref{wz2.12}) corresponds to $\Psi=F=G=1$. Other solutions of (\ref{wz2.16}) were provided by choosing $G=0$ what leads to the formula $\Psi(\xi)=\frac{d}{d\xi}F^2(\xi)$; in particular one gets \cite{32}
\begin{equation}
[\hat{x}_\mu,p_\nu]=i\eta_{\mu\nu}\sqrt{1-\beta(p^2+m^2)} \label{wz2.17}.
\end{equation}
It should be added that recently the Snyder models with algebraically independent operators $M_{\mu\nu}$ and $(\hat{x}_\mu,p_\mu)$ are called the extended Snyder models \cite{32},\cite{33}.

It should be added that the quantum phase spaces which similarly like in Yang models are described by noncommutative fourmomenta $\hat{p}$ (see Sect. 2.2) were introduced as algebraic characterization of so-called Triple Special Relativity models \cite{34},\cite{35}.

\section{Supersymmetric Snyder models and quantum superspaces}
\subsection{From quantum spaces to quantum superspaces} 

It is known that in Snyder type models the quantum-deformed spaces are described by Lie-algebraic coset generators. In particular in $D=4$ relativistic case there are chosen $\frac{O(4,1)}{O(3,1)}$ (dS) and $\frac{O(3,2)}{O(3,1)}$ (AdS) coset generators, which provide irreducible NC vectorial representations of $D=4$ Lorentz algebra describing quantum space-times.

One can generalize in natural way such constructions to Lie superalgebras \cite{22} with supercoset generators describing quantum-deformed superspaces. The relativistic quantum superspaces are the particular reducible graded NC representations of Lorentz algebra, with bosonic (vectorial) even and fermionic (spinorial) odd sectors. In special cases of supercosets $\frac{OSp(1,2k|R)}{Sp(2k)}$ and $\frac{OSp(1,2k|C)}{Sp(2k;C)}$ we get the irreducible quantum-deformed fermionic (Grassmanian) spinors: in real case one obtains only quantum-deformed $D=3$ Lorentz spinors $(k=1)$ and $D=4$ AdS spinors $(k=2)$; in complex case for $(k=1)$ one gets only $D=4$ complex quantum-deformed Weyl spinors.\footnote{If we choose quaternionic coset $\frac{OSp(1,2k|H)}{O(1|H)\otimes Sp(2k|H)}$ the $k=1$ case provides the quantum-deformed NC 2-component $\overline{O(4,1)}=Sp(2|H)$ quaternionic Weyl spinors, with additional  for $O(1|H)\simeq O(1,1)$ group describing internal symmetry (see \cite{36},\cite{37},\cite{22}).} Other example of spinorial coset is described by the generators of the coset $\frac{SU(2,2|1)}{U(2,2)}$, describing NC $D=4$ conformal spinors which can be identified as fermionic quantum-deformed $D=4$ twistors \cite{38}.

In following subsection of Sect. 3 we describe by supersymmetrization of AdS Snyder model the quantum $D=4$ superspace (Sect. 3.2) and from $D=4$ dS superalgebra (Sect. 3.3) one gets the quantum $D=4$ dS superspaces. In Sect. 4 we will discuss quantum phase spaces obtained by supersymmetrization of $D=4$ AdS and $D=4$ dS quantum-deformed phase spaces described by Yang models in Sect. 2.2. It should be added that the supersymmetrization of $D=4$ Snyder quantum phase spaces, considered in Sect. 2.3, is an interesting task which should be studied by choosing proper postulates for the supersymmetric counterpart of the phase space variables $(\hat{x}_\mu,p_\mu).$

\subsection{Supersymmetric $D=4$ Snyder models describing $D=4$ quantum AdS superspaces}
In $osp(1|4)$ superalgebra the generators $M_{\mu\nu}$ and $\x_\mu=\lambda M_{\mu 4}$ form the $D=4$ AdS $o(2,3)\simeq sp(4)$ subalgebra, with the generators $M_{\mu\nu}$ describing its $D=4$ Lorentz subalgebra. The $osp(1|4)$ superalgebra is obtained by adding to $D=4$ AdS algebra the four additional real Majorana supercharges $Q_\alpha$ which are interpreted as quantum NC fermionic real components of D=4 AdS spinors
   $\hat{\xi}_\alpha$. The $osp(1|4)$ superalgebra is described by the following supersymmetric extension of relations (\ref{wz0})-(\ref{wz2}) with $\epsilon=-1$ with the dimensionality $[\beta]=L^2$ 
\begin{eqnarray}
 [M_{\mu\nu},M_{\rho\sigma}] &= &
i(\eta_{\mu\sigma}M_{\nu\rho} + \eta_{\nu\rho}M_{\mu\sigma} 
- \eta_{\mu\rho}M_{\nu\sigma} -\eta_{\nu\sigma}M_{\mu\rho})\label{1lineads}\\
\cr
[M_{\mu\nu},\x_{\rho}]&=&i(\eta_{\nu\rho}\x_{\mu}-\eta_{\mu\rho}\x_{\nu})\label{lorr1}\\
\cr 
[\x_{\mu},\x_{\nu}]&=&-i\beta M_{\mu\nu}\\
\cr
 \left\{{\xih}_{\alpha}, \xih_{\beta}\right\}
&= & - 2(C\gamma^{\mu})_{\alpha\beta}\x_{\mu}
  + \beta^{\frac{1}{2}} \left(C\gamma^{\mu\nu}\right)_{\alpha\beta}
M_{\mu\nu}\label{qq}\\
\cr  
 [M_{\mu\nu},\xih_{\alpha}] &=&- \frac{i}{2} \xih_{\beta}
\left(\gamma_{\mu\nu}\right)^{\beta}_{\ \alpha}\label{lorr2}
\\ \cr
 [\x_{\mu},\xih_{\alpha}] &=&-
 \frac{i}{2}\beta^{\frac{1}{2}} \xih_{\beta}
\left(\gamma_{\mu}\right)^{\beta}_{\ \alpha}.\label{6lineads}
\end{eqnarray}
where the quantum spinors $\hat{\xi}_\alpha$
 appearing in the place of supercharges 
 have (see(\ref{qq})) the length dimensionality $[\xih_\alpha]=L^{\frac{1}{2}}$. The parameter $\beta$ is the $AdS_4$ radius usually identified in QG applications with Planckian length square and $\gamma_\mu$ are $D=4$ Dirac $O(3,1)$ matrices in real Majorana representation;
$
\gamma_{\mu\nu}={\frac{1}{2}}(\gamma_\mu\gamma_\nu-\gamma_\nu\gamma_\mu)
$.
Further, by $C_{\alpha\beta}=(\gamma_0)_{\alpha\beta}$ we denote the charge conjugation matrix with the properties
$C^T=-C,~(\gamma^\mu C^{-1})^T=-\gamma^\mu C^{-1}, C^2=-1$

The superalgebra (\ref{1lineads})-(\ref{6lineads}) describes graded associative quantum superspace $(\hat{x}_\mu^{AdS};\hat{\xi}_\alpha)$
\begin{equation}
\mathbb{X}^{(4,4)}_{AdS}=(\hat{x}_\mu^{AdS};\hat{\xi}_\alpha|M_{\mu\nu})
\end{equation}
where six generators $M_{\mu\nu}$ describe the $D=4$ Lorentz covariance algebra (see (\ref{lorr1}) and (\ref{lorr2})). Alternatively, by considering the coset $\frac{OSp(1;4)}{Sp(4)}$ one can consider in $\hat{osp}(1;4)$ only the supercharges, and obtain in such a way purely spinorial model of Snyder type $(\hat{\xi}_\alpha^{AdS}|M_{AB})$, with anticommuting $D=4$ AdS quantized spinors $\hat{\xi}_\alpha^{AdS}$, which are covariant under the $D=4$ AdS $O(3,2)$ transformations.

\subsection{Quantum $D=4$ dS superspace from $D=4$ SUSY dS Snyder model}

For $D=4$ dS algebra $o(4,1)$ the supercharges should be described by fundamental spinor realizations of the quaternionic spinorial covering of $D=4$ dS group $\overline{O(4,1)}=U(1,1;H)\equiv OSp(1,2;H)$ \cite{36}-\cite{37}. The $N=1$ supersymmetrization of $D=4$ dS algebra requires a pair of quaternionic supercharges, which following isomorphism $u(1,1|H)=usp(2,2;C)$ can be equivalently represented by the pair of four-component complex spinors $\hat{Q}_A^i (i=1,2;A=1\dots 4)$ with their quaternionic structure represented by symplectic $SU(2)$ Majorana condition \cite{39}-\cite{40}.

The simple $(N=1)$ $D=4$ dS quaternionic superalgebra is described as the intersection of two complex superalgebras\footnote{We use the notation,
 where N denotes the number of 2-component irreducible quaternionic D=4 dS supercharges (see e.g. \cite{39}, \cite{40}).}
\begin{equation}
\label{3.8}
uu_\alpha(1,1;1|H)=su(2,2;2)\cap osp(4;2|C)
\end{equation}
with bosonic sector $u(1,1;H)\oplus u_\alpha(1;H)\equiv usp(2,2)\oplus o(2)$.
 Using complex spinors notation the superalgebra (\ref{3.8})
   is described by the following set of (anti)commutators $(A,B=0,1,2,3,4;~\alpha,\beta=1,2,3,4;~ i,j=1,2)$
\begin{eqnarray}
 [M_{AB},M_{CD}] &= &
i(\eta_{AD}M_{BC} + \eta_{BC}M_{AD} 
- \eta_{AC}M_{BD} -\eta_{BD}M_{AC})\label{11lin}\\
\cr
 \left\{\hat{Q}^i_{\alpha}, \hat{Q}^j_{\beta}\right\}
&= & \delta^{ij}(\Sigma_{AB} C)_{\alpha\beta}M^{AB}+\epsilon^{ij}C_{\alpha\beta}T\\
\cr  
 [M_{AB},\hat{Q}^i_{\alpha}] &=&- (\Sigma_{AB})_{\alpha\beta}\hat{Q}_{\beta}^i
\\ \cr
 [M_{AB},T] &=&0\\
 \cr
 [T,\hat{Q}^i_\alpha]&=&-\epsilon^{ij}\hat{Q}^j_\alpha
 \end{eqnarray}
where $\eta_{AB}=diag(-1,1,1,1,1)$, i.e $M_{AB}$ are $o(4;1)$ generators, $T$ is a scalar internal $o(2)$ symmetry 
 generator and $\Sigma_{AB}= \frac{1}{2}[ \gamma_A, \gamma_B ]$ represents the 4$\times$4 complex matrix realization of $o(4,1)$ algebra (\ref{11lin}).
The complex $o(4,1)$ Dirac matrices
      can be chosen for A=0,1,2,3 as real $(\gamma^{dS}_\mu = \gamma_\mu)$
      and for $A=4$ the choice $\gamma^{dS} _4 = i \gamma_5$ is purely imaginary.
     The fermionic supercharges $\hat{Q}^i_\alpha$ satisfy the following quaternionic 
     $SU(2)$-symplectic Majorana condition
\begin{equation}
\hat{Q}^i_\alpha=\epsilon^{ij}(\gamma_5\hat{\bar{Q}}^j)_\alpha,\qquad\Qb=\Q^\dagger C.
\label{jjk}
\end{equation}
We substitute the generators in the coset $\frac{UU_\alpha (1,1;1|H)}{Sl(2;\mathbb{C})}$
 by quantum D=4 dS  superspace coordinates  as follows
\begin{equation}
\hat{\psi}_\alpha=\sqrt{i}\beta^{-\frac{1}{4}}\hat{Q}^1_\alpha,\qquad
 \hat{\psi}^\dagger_\alpha=-\sqrt{i}\beta^{-\frac{1}{4}}(\gamma_5\hat{Q}^2)_\alpha,\qquad \x_\mu=\beta^{\frac{1}{2}}M_{\mu4},
\end{equation}
and we obtain 
 the following superalgebra defining quantum D=4 dS superspace
\begin{eqnarray}
 [M_{\mu\nu},M_{\rho\sigma}] &= &
i(\eta_{\mu\sigma}M_{\nu\rho} + \eta_{\nu\rho}M_{\mu\sigma} 
- \eta_{\mu\rho}M_{\nu\sigma} -\eta_{\nu\sigma}M_{\mu\rho})\\
\cr  
[M_{\mu\nu},\x_{\rho}]&=&i(\eta_{\nu\rho}\x_{\mu}-\eta_{\mu\rho}\x_{\nu})\\
\cr 
[\x_{\mu},\x_{\nu}]&=&i\beta M_{\mu\nu}\\
\cr 
 \left\{\hat{\psi}_{\alpha},\hat{ \psi}_{\beta}\right\}
&= & -i(\gamma^{\mu}C)_{\alpha\beta}\x_{\mu}
  +i\beta \left(\gamma^{\mu\nu}C\right)_{\psi\beta}
M_{\mu\nu}\label{srd}\\
\cr 
 \left\{\hat{\psi}^\star_{\alpha}, \hat{\psi}^\star_{\beta}\right\}
&= & - i(\gamma^{\mu}C)_{\alpha\beta}\x_{\mu}
  -i\beta\left(\gamma^{\mu\nu}C\right)_{\alpha\beta}
M_{\mu\nu}\\
\cr 
 \left\{\hat{\psi}_{\alpha}, \hat{\psi}^\star_{\beta}\right\}
&= & -(\gamma_5)_{\alpha\beta}T   \label{ghost}\\
\cr 
 [M_{\mu\nu},\hat{\psi}_{\alpha}] &=&-  
\left(\gamma_{\mu\nu}\right)^{\beta}_{\ \alpha}\hat{\psi}_{\beta}
\\ \cr 
 [\hat{x}_{\mu},\hat{\psi}_{\alpha}] &=&i\beta^\frac{1}{2}
\left(\gamma_{\mu}\right)^{\beta}_{\ \alpha}\hat{\psi}_{\beta}
\\ \cr 
 [T,\hat{\psi}_{\alpha}] &=&\gamma_5 \hat{\psi}^\star_{\alpha}\label{llas}
\end{eqnarray}
with the length dimensionalities
\begin{equation}
[M_{\mu\nu}]=0,\qquad[\x_\mu]=1,\qquad
[\hat{\psi}_\alpha]=[\hat{\psi}_\alpha]=\frac{1}{2},\quad[T]=1.
\end{equation}
The superalgebraic relations define the quantum $D=4$ dS superspace
\begin{equation}
\mathbb{X}_{dS}^{(5;4+\bar{4})}=(\hat{x}_\mu^{dS};\hat{\psi}_\alpha,\hat{\psi}_\alpha^\star|M_{\mu\nu},T)
\end{equation}
where $M_{\mu\nu}$ are $D=4$ Lorentz generators and $T$ describes the generator of internal $O(2)$ symmetries. 

It follows from relation (\ref{ghost}) and traceless $\gamma_5$ matrix that by putting $\alpha=\beta$ in (\ref{ghost}) one gets $\sum_{\alpha=1}^4|\hat{\psi}_\alpha|^2=0$. The nonvanishing quantum spinors can be therefore only realized in Hilbert-Krein space of states with indefinite metric (see e.g. \cite{41},\cite{42}) and local gauging of quaternionic superalgebra (\ref{3.8}) leads to $D=4$ dS supergravity which contains necessarily gauge ghost fields \cite{43}. 

\section{$N=2$ supersymmetric Yang models and quantum phase superspaces}

\subsection{Quantum $D=3$ Yang AdS phase superspace from $osp(2;4)$ superalgebra}
In order to obtain $D=3$ SUSY AdS Yang model we add to the $o(3,2)$ generators $M_{AB}$ (see (\ref{wz0}), $\epsilon=-1$) describing $D=3$ AdS Yang model the pair of real $O(3,2)$ spinorial supercharges $Q^i_\alpha$ $(i=1,2;\alpha=1\dots 4)$ and $o(2)$ internal symmetry generator $T$. The underlying $osp(2;4)$ superalgebra looks as follows (see e.g. \cite{44})
\begin{eqnarray}
&&\{Q_\alpha^i,Q_\beta^j\}=\delta^{ij}(\gamma^{AB}C)_{\alpha\beta}M_{AB}-\epsilon^{ij}C_{\alpha\beta}T\label{genaa}\\
&&[M_{AB},Q_\alpha^i]=-(\gamma_{AB})_\alpha^{~\beta}Q_\beta^i\\
&&[T,Q_\alpha^i]=-\epsilon^{ij}Q_\alpha^j\label{genaa1}
\end{eqnarray}
where $\gamma_{AB}=\frac{1}{2}[\gamma_A,\gamma_B]$ and $\gamma_A$ denotes the real $O(3,2)$ Dirac-Majorana matrices. Part of the generators in (\ref{genaa})-(\ref{genaa1}) describe the $D=3$ Lorentz covariance generators $M_{rs}(r,s=0,1,2)$, internal covariance generator $T$ and generator $M_{34}=o(1,1)$ interchanging coordinates and momenta
\begin{equation}
M_{rs}\oplus M_{34}\oplus T\simeq o(2,1)\oplus o(1,1)\oplus o(2).
\label{p4.4}
\end{equation}
The remaining generators are assigned to the $D=3$ AdS quantum-deformed superphase space coordinates as follows
\begin{equation}
M_{3r}\oplus M_{4r}\oplus \tilde{Q}^1_\alpha\oplus \tilde{Q}^2_\alpha\simeq \frac{1}{\lambda}\hat{x}_r\oplus R\hat{p}_r\oplus \hat{\xi}_\alpha\oplus\hat{\pi}_\alpha
\label{p4.5}
\end{equation}
where $(\hat{x}_r,\hat{p}_r)$ and $(\hat{\xi}_\alpha,\hat{\pi}_\alpha)$ describe even and odd canonical pairs of vectorial and spinorial positions and momenta. Introducing the generator $\hat{d}=\frac{\lambda}{R}M_{34}$ we obtain the following SUSY-extended quantum-deformed Heisenberg algebra
\begin{itemize}
\item[i)] odd-odd relations
\begin{eqnarray}
&&\{\hat{\xi}_\alpha, \hat{\xi}_\beta\}=\{\hat{\pi}_\alpha, \hat{\pi}_\beta\}=(\gamma^{rs}C)_{\alpha\beta}M_{rs}+\frac{1}{\lambda}(\gamma^{3r}C)_{\alpha\beta}\hat{x}_r+R(\gamma^{4r}C)_{\alpha\beta}\hat{p}_r+(\gamma^{34}C)_{\alpha\beta}\hat{d}\nonumber\\
&&\{\hat{\pi}_\alpha, \hat{\xi}_\beta\}=C_{\alpha\beta}T\label{a1}
\end{eqnarray}
\item[ii)] even-even relations describing deformed Heisenberg algebra
\begin{eqnarray}
&&[\hat{x}_r,\hat{x}_s]=\lambda^2M_{rs},\qquad [\hat{p}_r,\hat{p}_s]=\frac{1}{R^2}M_{rs}, \qquad [\hat{p}_r,\hat{x}_s]=i\eta_{rs}\hat{d} \nonumber
\end{eqnarray}
\item[iii)] crossed even-odd relations
\begin{eqnarray}
&[\hat{x}_r,\hat{\xi}_\alpha]=(\gamma_{3r})_\alpha^{~~\beta}\hat{\xi}_\beta \qquad &[\hat{p}_r,\hat{\xi}_\alpha]=(\gamma_{4r})_\alpha^{~~\beta}\hat{\xi}_\beta \nonumber\\
&[\hat{x}_r,\hat{\pi}_\alpha]=-(\gamma_{3r})_\alpha^{~~\beta}\hat{\pi}_\beta \qquad &[\hat{p}_r,\hat{\pi}_\alpha]=-(\gamma_{4r})_\alpha^{~~\beta}\hat{\pi}_\beta.\label{a3}
\end{eqnarray}
\end{itemize}
The relations above can be supplemented by $D=3$ version of relations (\ref{l11})-(\ref{l22}). We see that the relations (\ref{a1})-(\ref{a3}) depend on the quantum superspace coordinates (\ref{p4.5}) as well as the covariance symmetry generators (\ref{p4.4}). We add that using $6\times 6$ graded matrix representation  of $osp(2;4)$ one obtains the matrix realization of the $D=3$ phase superspace superalgebra.
\subsection{$D=4$ AdS quantum phase superspace $su(2,2;2)$}
Using $N=2$ superextension of $D=4$ Minkowskian conformal algebra 
\begin{equation}
o(4,2)\simeq su(2,2)\xrightarrow{\text{SUSY}}su(2,2;2)
\end{equation}
one can define by the coset $\frac{SU(2,2;2)}{SU(2,2)\times U(2)}$ generators the spinorial coordinates  and momenta which span the odd sector of relativistic $D=4$ supersymmetric Heisenberg algebra. The full covariance algebra is described by the following extension od $D=4$ Lorentz algebra $(\mu,\nu=0,1,2,3)$
\begin{equation}
M_{\mu\nu}\rightarrow (M_{\mu\nu}, M_{45}, I, I_r)\simeq o(3,1)\oplus o(1,1)\oplus o(2)\oplus o(3)\label{4.9}
\end{equation}
where $(I, I_r)$ $(r=1,2)$ describe the internal $u(2)$ symmetries commuting with conformal $o(4,2)$ algebra. The remaining generators introduce the quantum-deformed superspace coordinates of $D=4$ AdS phase superspace \footnote{We call the operation $\xrightarrow{S}$ the Snyderization procedure of Lie algebra generators (see \cite{22}).}:
\begin{equation}
(M_{3\mu}, M_{4\mu}, \tilde{Q}_\alpha^a, \tilde{S}_\alpha^a)\xrightarrow{S} (\frac{1}{\lambda}\hat{x}_\mu, R\hat{p}_\mu, \hat{\psi}^a_\alpha, \hat{\pi}^a_\alpha).\label{4.10}
\end{equation}

The fermionic sector of $su(2,2;2)$ can be conveniently  described by two pairs of four component real Majorana supercharges $Q^a_\alpha, S_\beta^a$ $(a=1,2;\alpha,\beta=1\dots 4)$, satisfying the following basic superalgebraic relations (we use standard notation $(M_{\mu\nu}, P_\mu, K_\mu, D)$ for the $D=4$ conformal algebra generators; see \cite{45},\cite{46})
\begin{eqnarray}
&&\{Q_\alpha^a,Q_\beta^b\}=2\delta^{ab}(\gamma^\mu C)_{\alpha\beta}P_\mu\nonumber\\
&&\{S_\alpha^a,S_\beta^b\}=-2\delta^{ab}(\gamma^\mu C)_{\alpha\beta}K_\mu\label{4.11}\\
&&\{Q_\alpha^a,S_\beta^b\}=\delta^{ab}[(\gamma^{\mu\nu} C)_{\alpha\beta}M_{\mu\nu}+2iC_{\alpha\beta}D] +\epsilon^{ab}C_{\alpha\beta}I_2+i\tau^{(ab)}_k(\gamma_5C)_{\alpha\beta}I_k\nonumber
\end{eqnarray}
where $k=0,1,3$, three $2\times 2$ symmetric matrices are $\tau^{(ab)}_k=(1_2,\sigma_1, \sigma_3)$ and generators $(I_0, I_1, I_2, I_3)$ describe internal symmetry algebra $u(2)\simeq o(2)\oplus o(3)$. In order to reexpress $P_\mu, K_\mu$ by generators $M_{3\mu}, M_{4\mu}$ we proceed with the Snyderization procedure using the following formulae
\begin{equation}
M_{4\mu}=\frac{1}{\sqrt{2}}(RP_\mu+\frac{1}{\lambda}K_\mu)\xrightarrow{S}\frac{1}{\sqrt{2}\lambda}\hat{x}_\mu
\label{zm1}
\end{equation}
\begin{equation}
M_{5\mu}=\frac{1}{\sqrt{2}}(RP_\mu-\frac{1}{\lambda}K_\mu)\xrightarrow{S}\frac{R}{\sqrt{2}}\hat{p}_\mu.
\end{equation}
The supercharges $\tilde{Q}^a_\alpha, \tilde{S}^a_\alpha$ employed in the Snyderization procedure (\ref{4.10}) are defined in terms of supercharges $Q^a_\alpha, S^a_\alpha$ (see (\ref{4.11})) as follows $([Q^a_\alpha]=[\tilde{Q}^a_\alpha])=-\frac{1}{2},[S^a_\alpha]=[\tilde{S}^a_\alpha]=-\frac{1}{2}$ are length dimensionalities)
\begin{equation}
\tilde{Q}^a_\alpha=\frac{1}{\sqrt{2}}(Q^a_\alpha+(\lambda R)^{-\frac{1}{2}}S^a_\alpha)\qquad \tilde{S}^a_\alpha=\frac{1}{\sqrt{2}}(S^a_\alpha-(\lambda R)^{\frac{1}{2}}Q^a_\alpha).
\label{zm3}
\end{equation}
The algebra (\ref{4.11}) after using relations (\ref{zm1})-(\ref{zm3}) can be rewritten in term of the supercharges (\ref{zm3}) in the following way
\begin{eqnarray}
&&\{\tilde{Q}_\alpha^a,\tilde{Q}_\beta^b\}=\delta^{ab}[(\gamma^\mu C)_{\alpha\beta}\frac{1}{R}M_{5\mu}+(\gamma^{\mu\nu}C)_{\alpha\beta}M_{\mu\nu}]+\epsilon^{ab}C_{\alpha\beta}I_2\label{ll1}\\
&&\{\tilde{S}_\alpha^a,\tilde{S}_\beta^b\}=-\delta^{ab}[(\gamma^\mu C)_{\alpha\beta}\lambda M_{4\mu}+(\gamma^{\mu\nu}C)_{\alpha\beta}M_{\mu\nu}]-\epsilon^{ab}C_{\alpha\beta}I_2\\
&&\{\tilde{Q}_\alpha^a,\tilde{S}_\beta^b\}=-\delta^{ab}C_{\alpha\beta}D+i\tau^{(ab)}_k(\gamma_5C)_{\alpha\beta}I_k.\label{ll3}
\end{eqnarray}
Using relations (\ref{4.10}) we obtain from algebra (\ref{ll1})-(\ref{ll3}) the following fermionic odd-odd sector of $D=4$ AdS quantum-deformed phase superspace
\begin{eqnarray}
&&\{\psi_\alpha^a,\psi_\beta^b\}=\delta^{ab}[(\gamma^\mu C)_{\alpha\beta}\hat{p}_\mu+(\gamma^{\mu\nu}C)_{\alpha\beta}M_{\mu\nu}]+\epsilon^{ab}C_{\alpha\beta}I_2\label{ll1a}\\
&&\{\pi_\alpha^a,\pi_\beta^b\}=-\delta^{ab}[(\gamma^\mu C)_{\alpha\beta}\hat{x}_\mu+(\gamma^{\mu\nu}C)_{\alpha\beta}M_{\mu\nu}]-\epsilon^{ab}C_{\alpha\beta}I_2\\
&&\{\pi_\alpha^a,\psi_\beta^b\}=\delta^{ab}C_{\alpha\beta}D+i\tau^{(ab)}_k(\gamma_5C)_{\alpha\beta}I_k.\label{ll3a}
\end{eqnarray}
We see that in the above relations besides the "bosonic" $(\hat{x}_\mu, \hat{p}_\mu)$ and "fermionic" $(\psi_\alpha^a,\pi_\alpha^a)$ phase space coordinates, enter as well all the generators of covariance algebra (\ref{4.9}) (we recall that $D=M_{45}$).
\subsection{$D=4$ dS superphase space from $D=4$ SUSY dS Yang model}
Following the supersymmetrization of $\overline{o(5,1)}\simeq sl(2;H)$ algebra, in order to obtain $D=4$ supersymmetric dS Yang model one should consider the following $N=1$ quaternionic superalgebra isomorphic to $N=2$ complex superalgebra
\begin{equation}
sl(2|H)\simeq su^\star(4)\xrightarrow{\text{SUSY}} sl(2;1|H)\simeq su^\star(4,2)
\end{equation}
with bosonic sector 
\begin{equation}
sl(2|H)\otimes gl(1|H)\simeq su^\star(4)\otimes u^\star(2)
\end{equation}
where $su^\star(4)\simeq o(5,1)$ and $u^\star(2)=o(2)\oplus o(2,1)$. We see that due to quaternionic structure for our $D=4$ SUSY Yang models we should employ in complex notation the complex superalgebra $su^\star(4,2)$.

The superalgebras $su^\star(4;2N)$ can be obtained by so called Weyl trick (see e.g. \cite{23}) from $\hat{su}(4;2N)$ superalgebra which supersymetrizes $\hat{su}(4)\simeq\hat{o}(6)$. For $\hat{su}(4;2N)$ superalgebra one can introduce the following $Z_4$-graded superalgebra containing covariance algebra $L_0$ and the coset generators:
\begin{equation}
\begin{array}{cccc}
L_0 & L_1 & L_2 & L_3 \\
USp(4)\oplus USp(2N) & Q^+ & \frac{SU(4)}{USp(4)}\oplus\frac{U(2N)}{USp(2N)} & Q^-
\end{array}
\label{gens}
\end{equation}
where $Q^+,Q^-$ are the generators of the odd coset $\frac{SU(4;2N)}{USp(4)\otimes USp(2N)}$
and
\begin{equation}
[L_r,L_s\}=L_{r+s}\quad r=0,1,2,3 \quad mod~4.\label{multi}
\end{equation}
One passes from $su(4;2N)$ to $su^\star(4;2N)$ by multiplication of the generators from sectors (\ref{gens}) according to the following compact formula
\begin{equation}
su(4;2N)\rightsquigarrow su^\star(4;2N)\leftrightarrow L_r\rightsquigarrow \exp(\frac{ir\pi}{2})L_r.
\end{equation} 
We choose for $D=4$ SUSY dS phase space the following covariance algebra (compare with (\ref{4.9}))
\begin{equation}
(M_{rs}\oplus M_{45}\oplus I\oplus\tilde{I}_r)\simeq o(3,1)\oplus o(2)\oplus o(2)\oplus o(2,1).
\label{internal}
\end{equation}
The remaining generators, which include all supercharges, define via Snyderization procedure the $D=4$ dS quantum-deformed phase superspace
\begin{equation}
(M_{4\mu},M_{5\mu}, z^a_\alpha, u^b_\beta)\xrightarrow{S}(\frac{1}{\lambda}\hat{x}_\mu, R\hat{p}_\mu,\chi^a_\alpha,\rho^a_\alpha)
\label{snyderization}
\end{equation}
where due to quaternionic $SU(2)$-symplectic Majorana condition (see (\ref{jjk})) it follows that all independent degrees of freedom are described by $z_\alpha\equiv z^1_\alpha,\bar{z}_\alpha\sim z^2_\alpha$ and $u_\alpha\equiv u^1_\alpha, \bar{u}_\alpha\sim u^2_\alpha$ and after Snyderization by $\chi_\alpha\equiv\chi_\alpha^1,\bar{\chi}_\alpha\sim\chi^2_\alpha$ and $\rho_\alpha\equiv\rho_\alpha^1,\bar{\rho}_\alpha\sim\rho^2_\alpha$.

The $D=5$ dS superalgebra is the same as $D=4$ Euclidean conformal superalgebra, which has been studied earlier in explicit form (see e.g. \cite{33}). The fermionic odd-odd sector $(A=B=0,1,2,3,4,5; \bar{\Sigma}_{AB}=\frac{1}{2}[\gamma_A,\gamma_B]C$, where $C$ is a charge conjugation matrix for $D=5+1$) is the following

\begin{eqnarray}
&&\{z_\alpha,\bar{u}_\beta\}=2[(\bar{\Sigma}_{AB}\gamma_5)_{\alpha\beta}M^{AB}+C_{\alpha\beta}(\tilde{I}_1+i\tilde{I}_2)]\label{ghosta}\\
&& \{z_\alpha,\bar{z}_\beta\}=\{u_\alpha,\bar{u}_\beta\}=0\label{ghostb}\\
&&\{z_\alpha,u_\beta\}=i(\gamma_5C)_{\alpha\beta}(I+\tilde{I}_3)\label{ghostc}
\end{eqnarray}
and $I\oplus \tilde{I}_r~(r=1,2,3)$ describe the internal symmetry $o(2)\oplus o(2,1)$.

After Snyderization given by formulae (\ref{snyderization}) we get the algebra of complex spinors $\chi_\alpha, \rho_\alpha$ and $\bar{\chi}_\alpha,\bar{\rho}_\alpha$ representing the odd sector of $D=4$ dS phase superspace. The relations (\ref{ghosta})-(\ref{ghostc}) depend on the covariance algebra generators (\ref{internal}) and bosonic phase space coordinates $\hat{x}_\mu,\hat{p}_\mu$ are described by generators $M_{\mu 4},M_{\mu 5}$.

It should be mentioned that putting $\alpha=\beta$ in relations (\ref{ghost}) one gets that (see \cite{47}) $\sum_\alpha|z_\alpha|^2=\sum_\alpha|u_\alpha|^2=0$ or equivalently $\sum_\alpha|\chi_\alpha|^2=\sum_\alpha|\rho_\alpha|^2=0$. One can conclude that after quantization, similarly as in Sect. 3.2, the local gauging of superalgebra $su^\star(4;2)$ leads to $D=5$ dS supergravity which contains ghost gauge fields.

\section{Final Remarks and Outlook}

The Snyder dS and AdS models of NC Lorentz-covariant quantum space-time coordinates $\hat{x}_\mu$ are described by the algebras \begin{equation}
  o_{x} (4,1)  \xrightarrow {S}
 \frac{1}{\lambda}{\hat{x}}^{dS}_{\mu}  \oplus M_{\mu\nu}\qquad  o_{x} (3,2) \xrightarrow{S}
  \frac{1}{\lambda} \hat{x}^{AdS}_{\mu} \oplus M_{\mu\nu}  \label{5.1}
 \end{equation}
 Performing semi-dual Born mapp $\hat{x}_\mu \leftrightarrow \hat{p}_\mu,~\lambda\leftrightarrow \frac{1}{R}$
 one gets analogous algebraic structure with Lorentz-covariant quantum NC four-momenta ${\hat{p}}_\mu$

\begin{equation}
o_{p}(4,1)  
 \xrightarrow{S} R\hat{p}_\mu^{dS}  \oplus M_{\mu\nu} \qquad o_{p}(3,2) 
 \xrightarrow{S} R\hat{p}_\mu^{dS}  \oplus M_{\mu\nu}    \label{5.2}
 \end{equation}

Subsequently in dS and AdS Yang models the pairs of algebras (\ref{5.1}), (\ref{5.2}) are embedded in D=6 pseudo-orthogonal quantum deformed Yang algebras
  $o_H (5,1)$ and $o_H (4,2)$, where subindex $H$ denotes that they contain the basic generators $\hat{x}_\mu,\hat{p}_\mu$ of quantum-deformed relativistic Heisenberg algebra. We have the following two diagrams denoting the
   chains of  subalgebras

  \begin{equation}
  \begin{array}{cc}
  \begin{array}{c}
  o_{H} (5,1)\ \ 
  \\[-10pt]
  {\begin{array}{c}
 {\rotatebox{-25}{$\Huge\bigcup$}}

  \qquad\qquad 
  \begin{array}{c}
  \phantom{x}
  \\[-4pt]
   {\rotatebox{35}{$\Huge\bigcup$}}
   \end{array}
   \end{array}
   }
  \\
  o_x (4,1)  \supset o(3,1) \subset o_p(4,1)
  \end{array}
   \quad
   \begin{array}{c}
  o_{H} (4,2)
  \\[-10pt]
  \begin{array}{c}
 {\rotatebox{-25}{$\Huge\bigcup$}} 
  \qquad\qquad
  \begin{array}{c} \phantom{x}
  \\[-4pt]
   {\rotatebox{30}{$\Huge\bigcup$}}
   \end{array}
   \end{array}
  \\
  o_x (3,2)  \supset o(3,1) \subset o_p(3,2)
  \end{array}
  \end{array}
  \label{5.5}
  \end{equation}
  which imply the relativistic covariance of considered Snyder and Yang models.
  
  In this paper we introduced in Sect. 3 the $N=1$ SUSY generalizations of $D=4$ de-Sitter and anti-de-Sitter algebras and in  Sect.4 the $N=2$ supersymmetric extensions of $o_H (5,1)$ and $o_H (4,2)$ algebras. In such a way we defined the pairs of $N=1$ and $N=2$ SUSY Snyder models in order to introduce for $N=1$ the $dS$ and $AdS$ quantum superspaces and for $N=2$ the corresponding Lorentz-covariant quantum deformed Heisenberg superalgebras.
      
      In the  outlook we would like to comment on some possible directions of future studies, namely:
   \begin{itemize}
   \item[i)]
   Snyder and Yang type models are constructed from algebraic structures of classical Lie algebras. Before further development of such ideas one can pose the following two questions:
   \begin{enumerate}
   \item
   Can one repeat the above construction for quantum-deformed Lie algebras, in particular for quantum $o(4,1)$ and $o(5,1)$ algebras with noncocommutative Hopf-algebraic coalgebra sector and Hopf subalgebra describing the Lorentz symmetry? For such purpose one can employ the results obtained by Ballesteros et all [48]. In such a way one can obtain quantum-deformed Snyder and Yang models.
   \item
   Snyder and Yang type models are based on the decomposition of Lie algebras into the covariance subalgebra and the symmetric coset generators describing NC space-times (Snyder case) and NC quantum phase spaces described by quantum-deformed Heisenberg algebras (Yang case). Because the tangent space at every point of symmetric coset space has the structure of Lie triple system the algebraic description of symmetric cosets is provided by ternary algebras (see e.g \cite{49},\cite{50}). Can be therefore formulated quantum-deformed Snyder and Yang models in terms of respective quantum-deformed ternary algebras?
   \end{enumerate}
   The answer to these two questions is now considered  by present authors.
   
   \item[ii)] If one uses superalgebras for obtaining via Snyderization the spinorial degrees of freedom, the spinors will appear necessarily as Grasmannian, what is desirable in the framework of QFT. One can however employ also the "bosonic" cosets of matrix groups which describe the spinorial covering of space-time symmetry groups - in such a case one obtains curved bosonic spinors. A good example is provided by the case of conformal Penrose twistors $(t_A\in\mathbb{C};A=1\cdots 4)$, which are described by the fundamental representations of $D=4$ conformal group $SU(2,2)$. Defining twistors by bosonic cosets of $SU(2,3)$ or as $N=1$ $D=4$ superconformal odd cosets one gets the following two different choices of twistors (see e.g. \cite{50},\cite{51})  
  
\begin{equation}
\begin{array}{ccc}
\text{Penrose twistors}~~t_A^{(B)} & & \text{Fermionic twistors}~~t_A^{(F)}\\
t_A^{(B)}=\frac{SU(2,3)}{SU(2,2)\otimes U(1)} & \leftrightarrow & t_A^{(F)} =\frac{SU(2,2;1)}{SU(2,2)\otimes U(1)}.
\end{array}
\end{equation}

\item[iii)] It is known since eighties \cite{43},\cite{47} that local gauging of $D=4$ quaternionic dS superalgebras leads necessarily to the appearance of field-theoretic models with gauge ghost fields (see e.g \cite{43}). Recently, however, did appear an interesting  proposal (see e.g. \cite{52}) that $D=4$ dS supergravity can be obtained without ghost fields from  spontaneously broken coset $\frac{O(4,2)}{O(4,1)}\simeq\frac{SU(2,2)}{USp(2,2)}$ in $D=4$ superconformal gravity. Further algebraic understanding of this idea is still desired.

   \end{itemize}

\section*{Acknowledgements}

The authors would like to thank Stjepan Meljanac and Salvatore Mignemi for valuable comments. J.L. and M.W. have been supported by Polish National Science Center, project 2017/27/B/ST2/01902.

\end{document}